\documentclass[12pt]{article}
\renewcommand{\vec}[1]{{\bf#1}}

\newcommand{\be}{
\begin{eqnarray}}
\newcommand{\ee}{\end{eqnarray}}

\begin{document}

\rightline{RUB-TP2-05/03}
\begin{center}
{\Large On photoexcitation of baryon antidecuplet}\\[0.5cm]

M. V. Polyakov$^{a,b}$,  A. Rathke$^{c,d}$
\\[0.3cm]

 \footnotesize\it $^a$ Petersburg
Nuclear Physics Institute, Gatchina, St. Petersburg 188350,
Russia\\
\footnotesize\it $^b$ Institute for
Theoretical Physics II, Ruhr University Bochum, Germany\\
\footnotesize\it $^c$ Fakult\"at f\"ur Physik, Universit\"at Freiburg,
Hermann-Herder-Str.\ 3, 79104 Freiburg, Germany\\
\footnotesize\it $^d$ Institut f\"ur Theoretische Physik, Universit\"at zu K\"oln,
Z\"ulpicher Str.\ 77, 50937 K\"oln, Germany.

\vspace*{1cm}
\end{center}
\begin{abstract} We show that the photoexcitation of the
baryon antidecuplet, suggested by the soliton classification of low-lying baryons,
is strongly suppressed on the proton target.
The process occurs mostly on the neutron target.
This qualitative prediction can be useful in identifying the non-exotic members of the
antidecuplet in the known baryon spectrum. We also analyze the interrelation between photocouplings
of various baryon multiplets in the soliton picture and in the nonrelativistic quark model.
\end{abstract}
\vspace{0.43cm}

\noindent
{\bf 1.}~The soliton picture of baryons suggests a certain classification scheme
for the low-lying baryons. In this scheme various baryons appear as rotational
excitations of the same classical object -- soliton. In the case of three light
flavours, the first two low-lying $SU_{ fl}(3)$ multiplets are the octet
and the decuplet, just the same as in the quark model and in reality. The
third rotational excitation is an
 {\em antidecuplet with spin $1/2$}. Probably the existence of the
 antidecuplet as the next $SU_{fl}(3)$ rotational excitation has been
 first pointed out at the ITEP Winter School (February, 1984), see Ref.
 \cite{DP}.  Other early references for the antidecuplet include
Refs.~\cite{Chemtob, Prasz2, Walliser}.

 In Fig.~1 we draw the $SU_{fl}(3)$ diagram (from Ref.~\cite{DPP10}) for the
 suggested antidecuplet in the $(T_3,Y)$ axes, indicating its naive
 quark content as well as the (octet baryon + octet
 meson) content. In addition to the lightest
 $Z^+$, there is an exotic quadruplet of $S=-2$ baryons (we call them
 $\Xi_{3/2}$). In Ref.~\cite{DPP10} the following mass formula for
 the members of the antidecuplet was obtained:

 \be
 M=\biggl[1890-Y\times 180\biggr]\ {\rm MeV}\, .
 \ee
 Note that this ``soliton" mass formula is, to some extent, counterintuitive from the point of
 view of the naive quark model. For instance, strange baryon ($Z^+$)
appears to be lighter than the baryon with the nucleon quantum numbers. Up
to now we were used to strange baryons being heavier than
non-strange ones in a given multiplet.  Also $Z^+$ having 4 light+$\bar
s$ quark content is about 540~MeV lighter than $\Xi_{3/2}^-$ with the quark
content 3 light+2 $s$ quarks. In the naive quark model one would expect
 the mass difference of about $\sim$150~MeV.

The essential assumption made in Ref.~\cite{DPP10} was the identification of the
P$_{11}$ resonance, $N(1710)$, with the nucleon-like member of the antidecuplet.
The calculated decay modes of $N(1710)$ were found to be in a reasonable
agreement with the existing data. Note, however, that the data were not good enough
to make a decisive conclusion.
At least it seems that the standard non-relativistic $SU(6)$
description of this state as a member of an octet, is in trouble with the
data: the antidecuplet idea fits better.
With the identification made in Ref.~\cite{DPP10},
the lightest exotic member of the antidecuplet is $Z^+$ ($S=+1, Q=+1,
 T=0$) predicted to have the mass around $1530$~MeV and the total width of
less than $15$~MeV. As it was  discovered in Ref.~\cite{DPP10} the
exotic $Z^+$ should be anomalously narrow due to the specific interplay
of the
soliton rotational correction to the meson-baryon couplings. In
particular it was shown in Ref.~\cite{DPP10} that all these
couplings tend to zero in the non-relativistic quark limit
of the soliton picture of baryons.
The anomalous narrowness of $Z^+$ can explain why
it escaped the thorough searches in the past in $KN$ scattering processes.
For references, see the latest PDG report
on $Z$ baryons in 1986 Review~\cite{PDG2}
summarizing 20 years of experimental activity
on $S{=}+1$ baryons. Also, see the latest partial wave analysis for
$K^+ N$ scattering in Ref.~\cite{Hyslop:cs}.

Recently the first evidence of a {\em narrow} $S=+1$ resonance in the mass region of
$1530$~MeV has been reported by the LEPS collaboration at SPring-8  \cite{nakano}
and by DIANA collaboration at ITEP \cite{dolgolenko}.
If confirmed, this discovery may lead to a considerable revision of the
quark model baryon spectroscopy as we have known it for the last forty years.

In the present paper we show that the photoproduction of the
antidecuplet excitation of the chiral soliton possess qualitative
features which can be used as a clear signal for its identification. In
particular, we show that the photoexcitation of the baryon antidecuplet,
suggested by the soliton classification of low-lying baryons, is strongly
suppressed on the proton target. It occurs mostly on the neutron
target.  \vspace{0.1cm}

\noindent
{\bf 2.}~In order to estimate the photoexcitation of the antidecuplet of baryons, we shall
exploit the idea that all low-lying baryons are rotational excitations of the same classical
object --  the soliton.
We start with the magnetic dipole coupling of the soft photon with the momentum $\vec q$
to the soliton in the chiral limit:

\be
j_k^{\rm e.m.}(\vec q)=
\left[
v_1 \ D_{Qi}^{(8)}\left(R\right)
 +v_2\sum_{\alpha,\beta=4}^7
d_{i\alpha\beta}
D_{Q\alpha}^{(8)}\left(R\right)
J_\beta  +
 \frac{ v_3}{\sqrt 3}\cdot
D_{Q8}^{(8)}\left(R\right)
J_i\right]\
 i\varepsilon_{ijk}\ q_j \,.
\label{photon-soliton}
\ee
This equation requires a detailed  explanation. It is written in the space of collective
rotational coordinates, $R\in SU(3)$, of the soliton. The corresponding
operators of the infinitesimal $SU(3)$ rotation are denoted as
$J_A$, while
$D^{(\mu)}_{\nu\nu'}(R)$ stands for the Wigner $SU(3)$ finite-rotation
matrices depending on the orientation matrix of the soliton. Eventually $v_i$ are constants
which are universal
for all baryon multiplets.
In order to obtain the physical coupling of the photon to
baryons and various transitions using Eq.~(\ref{photon-soliton}),
 one has to sandwich it between the physical rotational states:

 \be
 \int dR\ \psi^*_{B_2}(R)\ldots \psi_{B_1}(R)\, ,
 \ee
 where the rotation wave function  of a particular baryon, $\psi_{B}(R)$,
 is expressed in terms of Wigner
 functions
\be
 \psi_{B}(R)=\sqrt{\dim \;r} (-1)^{J_3-1/2} D^{(\bar
r)}_{Y,T,T_3;1,J,-J_3},
\label{wave-functions}
\ee
 where $r$ is an irreducible representation of the
 $SU(3)$ group, $r=8,10,\overline{10}$, etc., $B$ denotes a set of quantum
 numbers $Y,T,T_3$ (hypercharge, isospin and its projection) and $J,
 J_3$ (spin and its projection). It is a big advantage of the
 chiral soliton picture that all concrete numbers (for masses and
 couplings) do not rely upon a specific dynamical realization but follow
 from symmetry considerations only.

An example of the usage of Eq.~(\ref{photon-soliton}) is the
calculation of the magnetic
moments of the octet and decuplet baryons in the chiral limit, see e.g.
Ref.~\cite{Magnetic moments,Kim}:

\noindent
\underline{Octet}
\be
\label{mm8u}
\mu_N&=&-\frac{14 T_3+1}{30}\ \left(v_1-\frac 12 v_2\right)+\frac{2T_3+3}{60}\ v_3 ,\\
\nonumber
\mu_\Sigma&=&-\frac{5 T_3+3}{30}\ \left(v_1-\frac 12 v_2\right)+\frac{5T_3-1}{60}\ v_3 ,\\
\nonumber
\mu_{\Sigma^0\Lambda}&=&-\frac{\sqrt 3}{10}\ \left(v_1-\frac 12 v_2+\frac 16 v_3\right) ,\\
\nonumber
\mu_\Lambda&=&\frac{1}{20}\ \left(v_1-\frac 12 v_2\right)+\frac{1}{120}\ v_3 ,\\
\label{mm8k}
\mu_\Xi&=&\frac{2T_3+2}{15}\ \left(v_1-\frac 12 v_2\right)+\frac{4T_3-1}{30}\ v_3 \,.
\ee

\noindent
\underline{Decuplet}
\be
\label{mm10}
\mu_B=-\frac{1}{8}\left(v_1-\frac 12 v_2-\frac 12 v_3\right)\ Q_B\, .
\ee
We give this example for two reasons. Firstly, it illustrates that the value of certain
combinations of
the universal constants
$v_i$ can be obtained from the data for the octet
and decuplet baryons magnetic moments (see detailes in Ref.~\cite{Magnetic moments}).
Secondly, using Eqs.~(\ref{mm8u}-\ref{mm10}) we can consider important limiting case of
the nonrelativistic quark model, which, to some extent, can be used as
a useful guiding line. In the nonrelativistic limit of the chiral
quark soliton model\footnote{For reviews of this model see
Refs.~\cite{DPP,Wakamatsu:1990ud,yellowbook,Dreview} and for review of close in spirit
NJL model see Refs.~\cite{tub,habweigel}.}
for the constants $v_i$, one obtains the following values\footnote{See
Refs.~\cite{DPP10,Praszalowicz:1995vi} for the discussion of the
nonrelativistic quark model limit in the soliton picture.}

\be
\label{NRL}
 v_2^{NR}/v_1^{NR}=-4/5, \qquad
v_3^{NR}/v_1^{NR}=-2/5.
\ee
Substituting these values into
eqs.~(\ref{mm8u}-\ref{mm8k}) and (\ref{mm10}) one obtains the famous
expressions for the magnetic moments of baryons in the
nonrelativistic quark model.
Calculations of $v_i$ in the chiral quark soliton model \cite{Kim}
confirm the negative sign of $v_{2,3}/v_1$
and give the following values:

\be
\frac{v_2}{v_1-\frac 12 v_2}= -0.3 \pm 0.08\ \ \left(-\frac 47\ {\rm NRL}\right) \qquad
\frac{v_3}{v_1-\frac 12 v_2}= -0.22 \pm 0.07\ \ \left(-\frac 27\ {\rm NRL}\right) \; ,
 \label{numest}
\ee
which indicate a deviation from the nonrelativistic quark model
results shown in the parentheses.  We note that the second
ratio is closer to the nonrelativistic quark model result. This ratio
is related to the strange magnetic moment of the nucleon \cite{Kim},
the nonrelativistic limit corresponding to $\mu_N^{s}=0$. Below for
our numerical calculations we shall use for the second ratio in
eq.~(\ref{numest}) its nonrelativistic value of $-2/7$.

Due to its universality eq.~(\ref{photon-soliton}) can be used for
computing various phototransition amplitudes between different baryon
multiplets. Let us first give, as an illustration, the corresponding
expressions for the transition magnetic moments between the octet and
decuplet baryons. For the octet-decuplet dipole magnetic transition we
obtain:
\be
\label{8to10u} \mu_{N\Delta}&=&-2 T_3\
\sqrt{\frac{2}{15}}\ \left(v_1-\frac 12 v_2 \right)\, ,\\
\mu_{\Sigma\Sigma^*}&=&- (T_3+1)\ \frac{1}{ \sqrt{30}}\ \left(v_1-\frac 12
v_2\right)\, ,\\
\mu_{\Lambda\Sigma^*}&=&- \ \frac{1}{\sqrt{10}}\ \left(v_1-\frac 12
v_2\right)\, ,\\
\mu_{\Xi\Xi^*}&=& \left(T_3+\frac 1 2\right)\ \frac{1}{ \sqrt{30}}\ \left(v_1-\frac 12
v_2\right)\, .
\ee
We have seen previously that using the values of the constants $v_i$
(\ref{NRL}) obtained by nonrelativistic limit of the quark soliton
model we reproduce $SU(6)$ relations for the magnetic moments.  This
illustrates that the ``soliton relations" reproduce successfully the
results of the $SU(6)$ quark model for the baryon magnetic moments.
However, if we now apply the nonrelativitic limit to the octet-decuplet
magnetic transitions, we obtain a deviation of ``soliton relations" from
those of the $SU(6)$ quark model:

\be
\mu_{N\Delta}^{\rm NR}=\frac{7}{\sqrt{30}} \ \mu_p^{\rm NR}\, ,
\ee
which should be contrasted with the $SU(6)$ relation \cite{su6}:
\be
\mu_{N\Delta}^{\rm SU(6)}=\frac{2}{3}\sqrt{2} \ \mu_p^{\rm SU(6)}\,.
\ee
Note that the ``soliton relations" for the octet-decuplet transitions,
even in the nonrelativistic limit,  are in better agreement
with experimental value of $\mu_{N\Delta}/\mu_p=1.24\pm 0.01$
\cite{Tiator:2000iy}
than the corresponding $SU(6)$ relations.

Now it is easy to derive the expressions for the dipole magnetic
transitions between the octet and antidecuplet baryons. The result is:

\be
\label{8toanti10u}
\mu_{N N^*}&=&-(2 T_3-1)\ \frac{1}{12\sqrt{5}} \left(v_1+v_2+\frac 12 v_3
\right)\, ,\\
\label{8toanti10d}
\mu_{\Sigma\Sigma^*}&=&- (T_3-1)\ \frac{1}{12\sqrt{5}}\ \left(v_1+v_2+\frac 12
v_3\right)\, .
\ee
We see immediately the important qualitative feature of the
octet-antidecuplet electromagnetic transitions:  in the chiral
limit the photoexcitation of the antidecuplet from the proton
[$T_3=1/2$ in Eq.~(\ref{8toanti10u})]
or
$\Sigma^+$ [$T_3=1$ in Eq.~(\ref{8toanti10d})] targets {\em does not
occur}. In our scheme the excitation of the anti-10 from the proton and
from $\Sigma^+$
 can occur only due to the $SU_{fl}(3)$ symmetry breaking
effects. Hence, the corresponding couplings should be relatively
suppressed.
This qualitative feature can be used experimentally as a test
whether a given $P_{11}$ nucleon resonance is a member of the
antidecuplet.  Another important feature of
Eq.~(\ref{8toanti10u}) is that in the nonrelativistic limit (see
(\ref{NRL})), the combination of constants $\left(v_1+v_2+\frac 12
v_3\right)$ is exactly zero. It means that the photoexcitation of the
antidecuplet
{\bf[} even if allowed {\bf ]}
is a purely relativistic
effect from the point of view of the quark model. This is also true
for the
meson decays of the antidecuplet. In particular this feature explains
why the exotic member $Z^+$ should be anomalously narrow, see
discussion in \cite{DPP10}.
\vspace{0.1cm}

\noindent
{\bf 3.}~In Ref.~\cite{DPP10} the nucleon-like
member of the antidecuplet has been identified with the nucleon resonance $P_{11}(1710)$.
Now using the results of Eq.~(\ref{8toanti10u}), we can make a prediction for the photon
dipole magnetic couplings for this resonance. To this end we have to fix the values
of the dynamical constants $v_i$.
We fix the ratio  $v_2/(v_1- v_2/2)$
to the value obtained in the chiral quark soliton model, see the first equation in (\ref{numest}),
whereas the ratio $v_3/(v_1-v_2/2)$ we fix by its nonrelativistic value of $-2/7$. The later
corresponds to the vanishing strange magnetic moment of the
nucleon. With this all constants but $(v_1-v_2/2)$  are fixed.
The constant $(v_1-v_2/2)$ is adjusted in order to reproduce  the magnetic moment of the
proton.  To estimate the $SU_{fl}(3)$ breaking effects (expected at the level of 15-20\%) due
to the nonzero strange quark mass we use the method and the results of
Refs.~\cite{Magnetic moments,Kim}.

With such fixing of the constants $v_i$ we obtain the following
range for dipole magnetic transition between the octet and
antidecuplet nucleons [in nuclear magneton]:

\be
\label{10anti10num}
\mu_{pp^*}=-0.15\div 0.15,\qquad \mu_{nn^*}=-1 \div -0.3\,.
\ee
We should note here that the obtained numerical values are very
sensitive to the values of the constants $v_i$: it is reflected in a
rather wide spread of our numerical predictions. These spreads were obtained
varying the value of $v_2$ and the values of the symmetry breaking
effects.  The most important conclusion we can make from the above values is
that the photoexcitation of $P_{11}(1710)$ as a member of antidecuplet
is favoured from the neutron target, since for ratio of the {\em
octet-antidecuplet}
dipole magnetic transition one expects that $\left|\frac{\mu_{nn^*}}{\mu_{pp^*}}
\right|>2$. The corresponding ratio for the {\em octet-octet}
transition is estimated as $\sim -2/3$.
The magnetic couplings of the
antidecuplet are rather small because these couplings are non-zero
owing to the relativistic effects only. In the nonrelativistic quark
model limit they would be exactly zero.

Let us note that we should keep in mind
important caveats in the above estimates. Firstly, the magnetic transitions were computed
in the soft photon limit. The actual
energy of the photon in the rest frame of $N(1710)$ is rather large, about
$600$~MeV, which can be hardly considered as soft.
This can lead to rather sizable corrections to the numerical
estimates (\ref{10anti10num}). However, these correction will not
change the ratio of the proton to neutron transition, meaning that
these corrections do not change the qualitative feature of dominance
of photoexcitation from the the neutron target.

Secondly, we have made our estimates assuming that
$P_{11}(1710)$ is a purely antidecuplet state. However, quantum
numbers of this state do not preclude  its mixing with the
corresponding states from the octet family.
The mixing can be strong if
the nucleon excitation belonging to the octet is close in mass to
$P_{11}(1710)$.
A pattern of such a mixing
has been considered in Ref.~\cite{radial} in the framework of a
particular variant of the Skyrme model \cite{rad}.
The model of Ref.~\cite{radial} gives
strong mixing pattern between almost degenerate
nucleon states from antidecuplet and from the octet.
It should be possible to verify this experimentally by accurate
measurements of properties of the nucleon
resonances in the mass region around 1700~MeV.
Unfortunately, present information about nucleon resonances
in this mass region is rather incomplete and controversial, see
the examples of recent analysises
\cite{Arndt:2002xv,Penner:2002md,Janssen:2003zv,Chiang:2001pw}.
For instance, in recent analysis of pion photoproduction data of
Ref.~\cite{Arndt:2002xv} the resonance $P_{11}(1710)$ is very
elusive, similar feature has been found in
Ref.~\cite{Janssen:2003zv} in the analysis of $\gamma p\to K^+\Lambda$
data.
Hopefully modern electron facilities like SPring-8
\cite{SP8}, Jlab \cite{cebafn}, ELSA \cite{ELSE}, MAMI \cite{Walcher:hy}, GRAAL
\cite{GRAAL} will bring us more detailed information on the nucleon resonances
in the 1700~MeV region.
\vspace{0.1cm}

\noindent
{\bf 4.}~In this paper we argued  that
the photoexcitation amplitudes are good tools
to probe the antidecuplet component of the nucleon resonances.
Probably special attention should be paid to the
{\em antidecuplet ``friendly"  photo-reactions} such as, for example,
\be
\gamma n\to K^+ \Sigma^-, \quad \gamma n\to \eta n, \quad \gamma n\to (\pi\pi)_{I=1} N\, .
\ee
In these channels the antidecuplet part of the nucleon resonances should be especially
enhanced, whereas in the analogous channels with the proton target the anti-10
component is relatively suppressed.  The anti-10 component can be also filtered out
in
{\em octet ``friendly" photo-reactions} such as
\be
\gamma p\to \gamma p, \quad \gamma p\to \pi \Delta, \quad \gamma p\to (\pi\pi)_{I=0}\ p.
\ee
High energy experiments can be also effectively used to search baryons from anti-10
family (pentaquarks),
see a detailed review on this \cite{Landsberg:rx}.

In the nonrelativistic limit of the ``soliton relations" for
photo- and meson\footnote{See detailed
discussion of meson couplings in Ref.~\cite{DPP10}.} couplings
of anti-10 baryons to ground state baryon octet we have found that they
would be zero in the nonrelativistic quark model. This important
qualitative feature makes anti-10 baryons (especially purely exotic
$Z^+$) some kind of benchmark for relativistic quark interactions
in baryons.

\section*{\normalsize\bf Acknowledgements}
We are thankful to Ya.~Azimov, D.~Diakonov, K.~Goeke, I.~Strakovsky, and
M.~ Vanderhaeghen for many
valuable discussions. The
work of MVP is supported
by the Sofja Kovalevskaja Prize of the Alexander von Humboldt
Foundation, the Federal Ministry of Education and Research and the
Programme for Investment in the Future of German Government.
 A.R.\ is supported by the DFG Graduiertenkolleg
``Nonlinear Differential Equations''.


\end{document}